\documentclass{PoS}

\usepackage{bm}
\usepackage{amsmath}

\title{Longitudinal spin asymmetries in polarized deuteron DIS with spectator tagging}

\ShortTitle{Longitudinal spin asymmetries in polarized deuteron DIS with spectator tagging}

\author{\speaker{W. Cosyn}\\
        Department of physics and Astronomy, Ghent University, Proeftuinstraat 86, B9000 Ghent, Belgium\\
        E-mail: \email{wim.cosyn@ugent.be}}

\author{C. Weiss\\
        Theory Center, Jefferson Lab, Newport News, VA 23606, USA\\
        E-mail: \email{weiss@jlab.org}}

\abstract{Polarized electron-deuteron DIS with spectator proton tagging offers a way of
measuring the neutron spin structure functions with maximal theoretical control of nuclear effects.
We calculate the nuclear structure factors in the longitudinal double-spin asymmetries
using methods of light-front nuclear structure. A unique feature of the spin-1 system is that spin
asymmetries can be formed either relative to the cross section in all three spin states ($\lambda_d = \pm 1, 0$) 
or in the two maximum-spin states only ($\pm 1$, involving tensor polarization). We find that the two-state 
deuteron spin asymmetry at small spectator proton momenta permits accurate extraction of the 
neutron structure function $g_{1n}$. Such measurements could be performed at a future electron-ion 
collider (EIC) with polarized deuteron beams and suitable forward detectors.}

\FullConference{23rd International Spin Physics Symposium - SPIN2018 -\\
		10-14 September, 2018\\
		Ferrara, Italy}

\begin{document}

\section{Introduction}
Deep-inelastic electron scattering on the deuteron with detection of a proton in the nuclear fragmentation 
region (``spectator tagging''), $e + d \rightarrow  e' + p + X$, represents an essential tool for studies
of nucleon structure and its nuclear modifications. The measurement of the proton recoil momentum 
($\lesssim$ 100 MeV in the deuteron rest frame) controls the nuclear configuration during the high-energy
process and permits a differential theoretical treatment of nuclear effects. On-shell extrapolation in the 
recoil momentum allows one to extract the free neutron structure functions, eliminating nuclear initial-state
modifications and final-state interactions \cite{Sargsian:2005rm}. Spectator tagging in polarized 
electron-deuteron DIS could be used for precise measurements of the neutron spin structure functions, 
which are needed for the flavor separation of the polarized quark densities and the singlet-nonsinglet 
separation in QCD evolution. Such measurements would become possible at a future electron-ion 
collider with polarized deuteron beams and forward proton detectors \cite{Boer:2011fh}.
It is therefore necessary to develop the theoretical description of polarized tagged DIS in collider 
kinematics and study its experimental feasibility \cite{Guzey:2014jva,LD1506,Cosyn:2016oiq}. 

In these proceedings we report a theoretical calculation of the nuclear structure effects in the longitudinal 
double spin asymmetries in tagged DIS in collider kinematics and discuss their implications for neutron spin 
structure extraction; see Ref.~\cite{Frankfurt:1983qs} for an earlier study. The nuclear effects are expressed 
in the form of a ``depolarization factor'' 
depending on the proton recoil momentum, describing the effective neutron polarization in the given 
nuclear configuration. We consider the spin asymmetries formed with all three deuteron spin states 
($\lambda_d = \pm 1, 0$) and the two maximum-spin states only ($\pm 1$) and compare their dependence on
the proton recoil momentum. We find that the two-state asymmetry permits a straightforward extraction of the 
neutron structure function $g_{1n}$. Details will be provided in a forthcoming article \cite{CW}.

\section{Spin asymmetries in tagged DIS}
\begin{figure}[t]
\parbox[c]{.3\textwidth}
{\includegraphics[width=.27\textwidth]{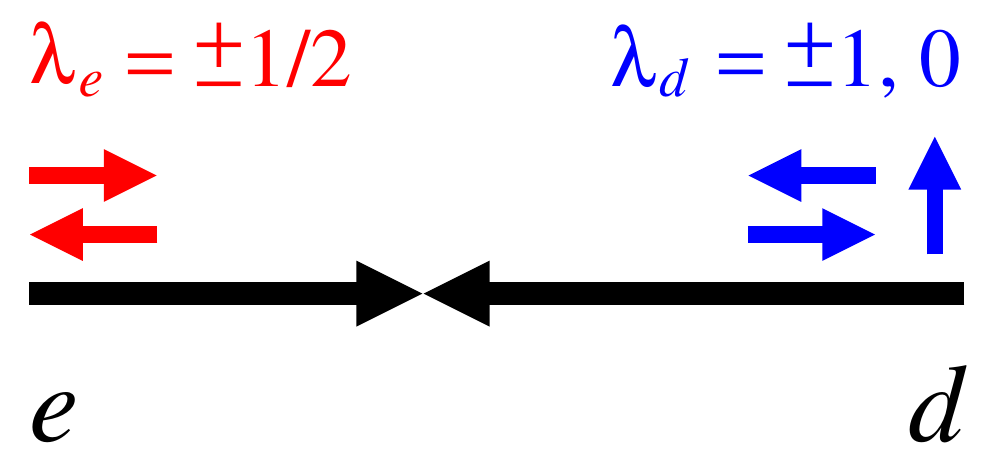}}
\hspace{.1\textwidth}
\parbox[c]{.5\textwidth}{\caption{\label{fig:electron_deuteron_spins}
Electron and deuteron helicity states in the colliding-beam setup}}
\end{figure}

Polarized tagged electron-deuteron DIS is described by the invariant differential cross section
\begin{equation}
d\sigma [ed \rightarrow e'pX] \; = \; {\mathcal F} dx \, dQ^2 \, d\Gamma_p[\textrm{recoil}], \hspace{2em}
{\mathcal F} \; = \; {\mathcal F}(x, Q^2; \textrm{recoil}; \lambda_e, \lambda_d),
\end{equation}
where $x$ and $Q^2$ are the usual DIS variables, ``recoil'' denotes the variables describing 
the proton recoil momentum (to be specified below), and $d\Gamma_p[\textrm{recoil}]$ is the invariant
phase space element in the recoil momentum. The variables $\lambda_e$ and $\lambda_d$ describe the electron 
and deuteron polarization in the initial state. In the following we consider a colliding-beam setup with 
collinear beams (zero crossing angle), in which $\lambda_e$ and $\lambda_d$ describe the spin projections
on the beam axis (see Fig.~\ref{fig:electron_deuteron_spins}); generalization to other situations is 
straightforward. Experiments typically measure 
differences and sums of cross sections in different electron and deuteron polarization states and 
their ratios (spin asymmetries). The ``polarized'' cross section is defined as the double-spin 
difference between the $\lambda_d = \pm 1$ deuteron spin states,\footnote{In fully inclusive DIS 
the spin dependence of the cross section is entirely through double-spin dependent terms 
$\propto \lambda_e \lambda_d$, and it is sufficient to take the difference in the electron or deuteron 
spin states alone, with the other spin remaining fixed. In tagged DIS the cross section can have 
also single-spin dependent terms $\propto \lambda_d$, and it is necessary to take the double difference 
in the electron and deuteron spins in order to isolate the double-spin dependent terms.}
%
%
\begin{equation}
d\sigma_\parallel \; \equiv \; 
{\textstyle\frac{1}{4}}
\left[ 
  \text{d}\sigma (+{\textstyle\frac{1}{2}}, +1)
- \text{d}\sigma (-{\textstyle\frac{1}{2}}, +1)
- \text{d}\sigma (+{\textstyle\frac{1}{2}}, -1)
+ \text{d}\sigma (-{\textstyle\frac{1}{2}}, -1)
\right] \,.
\label{sigma_polarized}
\end{equation}
It can be compared to two different ``unpolarized'' cross sections: (i) to the sum of all three deuteron spin
states ($\lambda_d = \pm 1, 0$); (ii) to the sum of the two maximum-spin states only ($\lambda_d = \pm 1$),
which enter in the polarized cross section Eq.~(\ref{sigma_polarized}),
\begin{eqnarray}
d\sigma^{(3)} &\equiv& \textstyle{\frac{1}{6}}
\left[ 
  \text{d}\sigma (+{\textstyle\frac{1}{2}}, +1)
+ \text{d}\sigma (-{\textstyle\frac{1}{2}}, +1)
+ \text{d}\sigma (+{\textstyle\frac{1}{2}}, -1)
+\text{d}\sigma (-{\textstyle\frac{1}{2}}, -1)
+ \text{d}\sigma (+{\textstyle\frac{1}{2}},  0)
+ \text{d}\sigma (-{\textstyle\frac{1}{2}},  0)
\right] ,
\nonumber \\
d\sigma^{(2)} &\equiv& \textstyle{\frac{1}{4}}
\left[ 
  \text{d}\sigma (+{\textstyle\frac{1}{2}}, +1)
+ \text{d}\sigma (-{\textstyle\frac{1}{2}}, +1)
+ \text{d}\sigma (+{\textstyle\frac{1}{2}}, -1)
+ \text{d}\sigma (-{\textstyle\frac{1}{2}}, -1)
\right] \, ;
\end{eqnarray}
the latter combination implies a nonzero tensor polarization of the deuteron ensemble.
Correspondingly, one can define the ``three-state'' and ``two-state'' spin asymmetries as
\begin{equation}
A_\parallel^{(3)} \equiv \frac{d\sigma_\parallel}{d\sigma^{(3)}}, 
\hspace{2em}
A_\parallel^{(2)} \equiv \frac{d\sigma_\parallel}{d\sigma^{(2)}}. 
\label{asymmetries}
\end{equation}
Both asymmetries can be used for neutron spin structure extraction from tagged DIS. They differ in the theoretical 
nuclear structure effects (S and D wave contributions, recoil momentum dependence) and the experimental requirements 
(preparation of $\lambda_d = 0$ spin state, systematics).
\section{Deuteron structure effects}
A theoretical calculation of the tagged DIS cross section has been performed using methods of light-front (LF) quantization, 
which permit the separation of nuclear and nucleonic structure in high-energy processes~\cite{Frankfurt:1983qs}. 
The description is implemented in a frame where the deuteron 4-momentum $p_d$ and the momentum transfer $q$ are 
collinear (along the $z$-axis), and the proton recoil momentum is described by its LF momentum variables
\begin{equation}
\alpha_p \; \equiv \; 2 p_p^+ / p_d^+ 
\hspace{1em}
(0 < \alpha_p < 2), \hspace{2em}
\bm{p}_{pT} .
\end{equation}
The deuteron's structure is described by a LF wave function in nucleon degrees of freedom 
(proton and neutron), which incorporates the relativistic spin structure (Melosh rotations connecting LF
helicity with canonical spin) and can be constructed approximately using the non-relativistic deuteron wave function 
in the rest frame (S and D waves). In the impulse approximation, the tagged spin asymmetries 
are obtained as (for cross sections integrated over the azimuthal angle of the recoil momentum) \cite{CW} 
\begin{equation}
A_\parallel^{(i)}(x, Q^2; \alpha_p, |\bm{p}_{pT}|) 
\; = \; D_d^{(i)}(\alpha_p, |\bm{p}_{pT}|) \;  
\frac{D_\parallel \, g_{1n}(\tilde{x},Q^2)}{2(1+\epsilon R)F_{1n}(\tilde{x},Q^2)}  
\hspace{2em} \textrm{for both $i = 2, 3$}.
\label{asymmetry_factorized}
\end{equation}
Here $g_{1n}$ is the neutron spin structure function, $\tilde{x}=x/(2-\alpha_p)$ is the effective scaling 
variable for the DIS process on the neutron, $R$ is the L/T ratio for the neutron, $\epsilon$ is the virtual 
photon polarization parameter, and 
\begin{equation}
D_\parallel = \frac{2y(1-y/2)}{1-y+y^2/2} 
\end{equation}
is the kinematic depolarization factor (the scaling variable $y$ is the same for DIS on the deuteron and
the free neutron up to small corrections). Eq.~(\ref{asymmetry_factorized}) applies in the
DIS limit $Q^2 \rightarrow \infty$, $x = Q^2/(p_d q)$ fixed; we neglect the power-suppressed contributions 
of the $g_{2n}$ structure function. The deuteron structure effects are contained in the factor $D_d^{(i)}$, 
which depends on the proton recoil momentum; it plays the role of a ``dynamical'' depolarization factor specific 
to the deuteron configuration selected by the proton recoil momentum. For the three-state asymmetry Eq.~(\ref{asymmetries})
the deuteron structure factor is obtained as
\begin{equation}
D_d^{(3)} (\alpha_p, |\bm{p}_{pT}|)
\; \equiv \;
\frac{\Delta f_d (\alpha_p, |\bm{p}_{pT}|)}{f_d (\alpha_p, |\bm{p}_{pT}|)}\,,
\end{equation}
where $\Delta f_d$ and $f_d$ are the helicity-dependent and -independent LF momentum distributions of 
neutrons in the deuteron, which are defined as certain densities of the deuteron LF wave function.
Their explicit expressions are \cite{CW}
\begin{align}
\Delta f_d(\alpha_p, |\bm{p}_{pT}|)  \; &= \; \frac{1}{2-\alpha_p}\left( U - \frac{W}{\sqrt{2}}\right) \left( U R_U
- \frac{W R_W}{\sqrt{2}} \right) \,, \nonumber\\
f_d(\alpha_p, |\bm{p}_{pT}|)  \; &= \; \frac{1}{2-\alpha_p} \left(U^2+W^2 \right)\,.
\end{align}
Here $U, W \equiv U(|\bm{k}|), W(|\bm{k}|)$ are the deuteron's radial wave functions in the S- and D-wave, and
\begin{align}
R_U(\bm{k}) 
\; &= \; 1 - \frac{(E + k^z) |\bm{k}_T|^2}{(E + m)(m^2 + |\bm{k}_T|^2)} \,,
\label{R_U_helicity_kt}
\\[2ex]
R_W(\bm{k}) 
\; &= \; 1 - \frac{(E + 2 m)(E + k^z) |\bm{k}_T|^2}{(m^2 + |\bm{k}_T|^2) |\bm{k}|^2} \,,	
\end{align}
are factors representing the relativistic spin effects; both depend on the effective center-of-mass
momentum of the proton-neutron system, $\bm{k}$, which is related to the LF momentum variables by
\begin{equation}
\alpha_p = 1 + \frac{k^z}{E}, \hspace{2em} \bm{k}_T = \bm{p}_{pT}, \hspace{2em} E =\sqrt{m^2+|\bm{k}|^2} ,
\end{equation}
and enables a rotationally symmetric representation of the two-body wave function~\cite{Frankfurt:1981mk,Kondratyuk:1983kq,Keister:1991sb}. 

For the two-state asymmetry Eq.~(\ref{asymmetries}) the deuteron structure factor is obtained as
\begin{equation}
D_d^{(2)} (\alpha_p, |\bm{p}_{pT}|)
\; \equiv \;
\frac{\Delta f_d (\alpha_p, |\bm{p}_{pT}|)}{f_d (\alpha_p, |\bm{p}_{pT}|) 
+ f_d(\alpha_p, |\bm{p}_{pT}|)[\text{tensor}]} ,
\end{equation}
where the additional contribution in the denominator is a tensor-polarized structure function.
It arises because the sum of $\pm 1$ spin states only (without the 0 state) corresponds to a
spin ensemble with an effective tensor polarization. Its explicit expression is \cite{CW}
\begin{equation}
f_d(\alpha_p, |\bm{p}_{pT}|)[\textrm{tensor}]
\; = \; - \frac{1}{2-\alpha_p} R_T\left( 2 U + \frac{W}{\sqrt{2}} \right) \frac{W}{\sqrt{2}}\,,
\end{equation}
with 
\begin{equation}
R_T(\bm k) = \left(1 - \frac{3 |\bm{k}_T|^2}{2 |\bm{k}|^2}\right) \,.
\end{equation}
One can show that the two-state factor satisfies the bounds
\begin{equation}
-1 \leq D_d^{(2)} \leq 1\, 
\label{eq:bound}
\end{equation}
because of the presence of the tensor-polarized term in the denominator;
no such bound is found for the three-state factor $D_3^{(3)}$, 
which can attain absolute values larger than unity.

\begin{figure}[t]
\includegraphics[width=.53\textwidth]{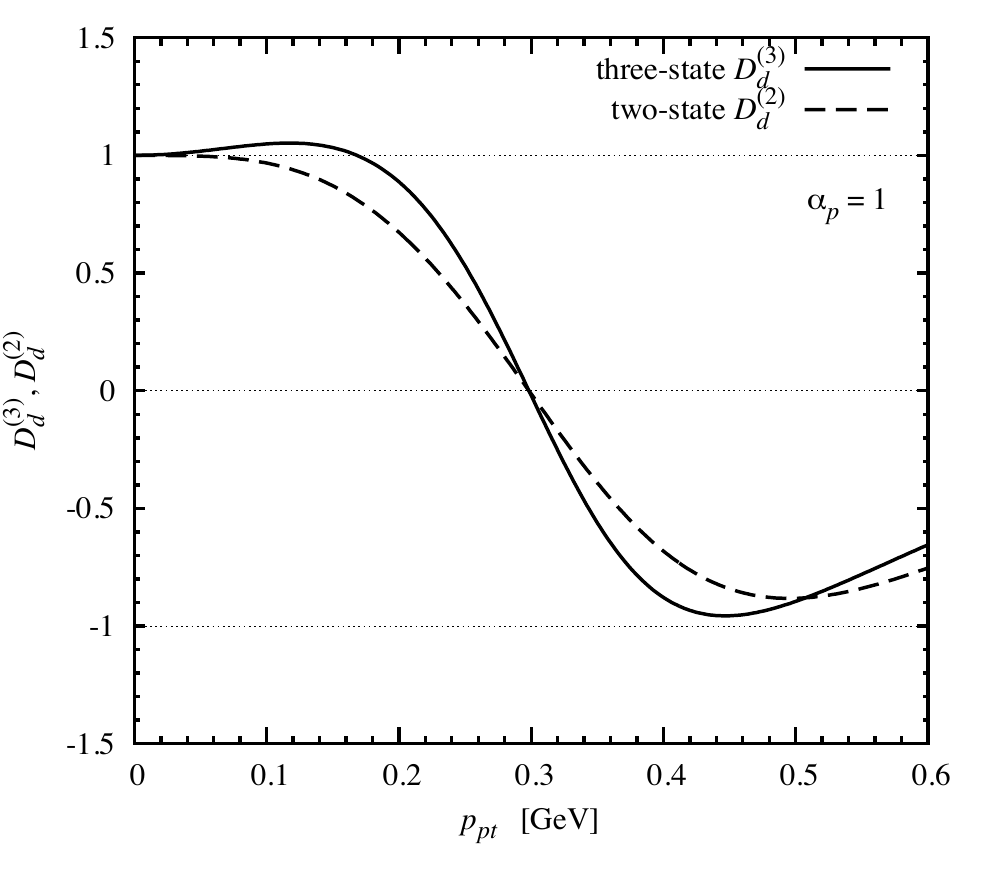}
\includegraphics[width=.47\textwidth]{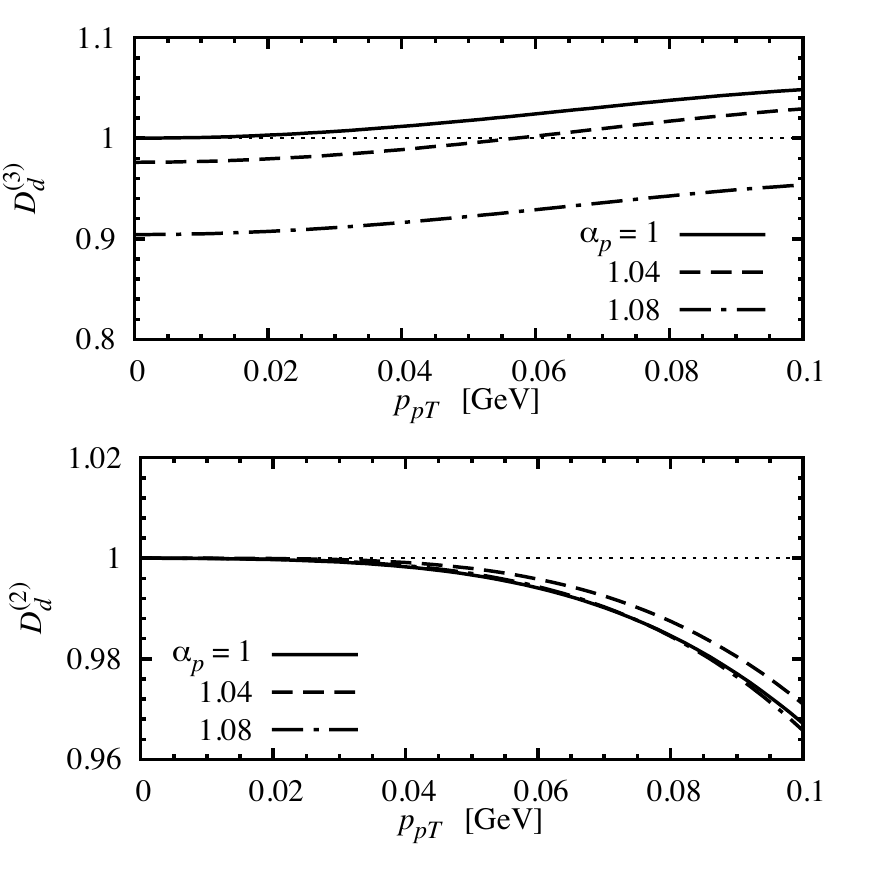}
\caption{Deuteron structure factors $D^{(3)}_d$ (three-state asymmetry) and $D^{(2)}_d$ (two-state asymmetry)
in polarized tagged DIS. Left panel: $D_d^{(3)}$ and $D_d^{(2)}$ at $\alpha_p = 1$ as 
functions of $p_{pT} \equiv |\bm{p}_{pT}|$. 
Right panels: $D_d^{(3)}$ and $D_d^{(2)}$ individually as functions of $|\bm{p}_{pT}|$, 
for several values of $\alpha_p$.}
\label{fig1}
\end{figure}

A numerical study of the deuteron structure factors has been performed using the AV18 
nonrelativistic wave function \cite{AV18} to construct the deuteron LF wave function 
and densities \cite{CW}. Figure~\ref{fig1} shows the two structure factors as a function
of $|\bm{p}_{pT}|$ and different values of $\alpha_p$. 
%
%
%
One observes:
\begin{enumerate}
\item Both $D_d^{(3)}$ and $D_d^{(2)}$ are unity at 
at $\alpha_p = 1$ and $|\bm{p}_{pT}| = 0$, where $|\bm{k}| = 0$ and only the S-wave is present. 

\item $D_d^{(3)}$ and $D_d^{(2)}$ 
remain close to unity for $|\bm{p}_{pT}| \lesssim$ 150 MeV, where the S-wave dominates. The D-wave 
contributions raise $D_d^{(3)}$ above unity but lowers $D_d^{(2)}$, in accordance with the  bound Eq.~(\ref{eq:bound}), showing the effect of the tensor polarized structure in $D_d^{(2)}$.

\item Both $D_d^{(3)}$ and $D_d^{(2)}$ decrease significantly at $|\bm{p}_{pT}| \gtrsim$ 150 MeV and
pass through zero at $|\bm{p}_{pT}| \approx$ 300 MeV, where the combination $(U - W/\sqrt{2})$ vanishes.
They become negative at larger momenta, where the D-wave dominates.

\item Because the factors $R_U, R_W$ and $R_T$ become unity at $\bm{k}_T = 0$,
$D_d^{(2)}= 1$ at $|\bm{p}_{pT}| = 0$ and arbitrary $\alpha_p$.
In contrast, $D_d^{(3)} \neq 1$ for $|\bm{p}|_{pT} = 0$ and arbitrary $\alpha_p$. Only if $\alpha_p = 1$, does $D_d^{(3)}$ become unity at $|\bm{p}|_{pT} = 0$.
\end{enumerate}

At small recoil momenta the values of $D_d^{(2)}$ are much closer to unity than those of 
$D_d^{(3)}$. This difference can also be seen in the expressions when one substitutes the factors
$R_U, R_W$ and $R_T$ by their approximate forms for small CM momenta $|\bm{k}| \ll m$.
In this approximation
\begin{align}
\left.
\begin{array}{lcl}
D_d^{(2)} \; &=& \; 1 \; + \; \textrm{terms} \; W^2/U^2
\\[1ex]
D_d^{(3)} \; &=& \; 1 \; + \; \textrm{terms} \; W/U
\end{array}
\right\} \hspace{2em} (|k| \ll m).
\end{align}
The D-wave corrections affect $D^{(2)}$ at only quadratic order at low momenta, but 
$D^{(3)}$ already at linear order.
\section{Neutron spin structure extraction}
In both the three-state and the two-state spin asymmetries the nuclear structure factors
approach unity at small proton recoil momenta (in the deuteron rest frame). The two-state
asymmetry has the advantage that the structure factor is identically equal to unity for
$\bm{p}_{pT}=0$ and any $\alpha_p$, resulting in a flat behavior of the spin asymmetry in 
the region where on-shell extrapolation will be performed. This feature should be convenient
for collider experiments, where longitudinal and transverse recoil momenta are measured
with different coverage and resolution (as determined by the beam optics and forward 
detector design). The two-state asymmetry also has the advantage that only preparation
of the $\pm 1$ deuteron spin states is required, reducing the systematic uncertainty.
The two-state asymmetry therefore appears to be the most promising method for extracting 
neutron spin structure from polarized tagged DIS experiments. It is interesting that the
asymmetry that is theoretically more complex (because of the presence of tensor polarization)
is practically more convenient for neutron structure extraction.

A full assessment of the prospects for neutron spin structure extraction also requires
an estimate of final-state interactions (FSI) between the tagged proton and the DIS products.
A detailed study has shown that FSI effects in unpolarized tagged DIS in EIC kinematics
are moderate, and that the on-shell extrapolation is still feasible~\cite{Strikman:2017koc}. 
The formalism presented there can be extended to the polarized case without essential changes.
\section{Applications and extensions}
The theoretical methods for high-energy scatering on the polarized deuteron with spectator tagging
described here can be applied and extended to several other types of measurements of interest \cite{CW}. 
This includes tagged DIS with transverse deuteron polarization; azimuthal asymmetries
and tensor-polarized tagged structure functions; tagged measurements of hard exclusive processes 
on the neutron such as deeply-virtual Compton scattering; and the use of tagging for studies
of nuclear modifications of partonic structure.

This material is based upon work supported by the U.S.~Department of Energy, 
Office of Science, Office of Nuclear Physics under contract DE-AC05-06OR23177.
\bibliographystyle{JHEP}
\bibliography{../bibtexall.bib}
\end{document}